\documentclass[a4paper, twocolumn, aps, prx, superscriptaddress, nofootinbib]{revtex4-1}
\usepackage{xcolor} 

\usepackage{hyperref}
\hypersetup{
	colorlinks,
	linkcolor={blue!90!black},
	citecolor={blue!10!blue},
	urlcolor={blue!80!black}
}
\usepackage{tikz}
\usepackage{amssymb}
\usepackage{amsmath}
\usepackage{graphicx}
\usepackage{multirow}

\usepackage{soul}
\usepackage[normalem]{ulem}

\definecolor{cobalt_blue}{rgb}{0, 0.2784313725, 0.7705882353}

\newcommand{\tauon}{\mathrm{T_{on}}}
\newcommand{\tauoff}{\mathrm{T_{off}}}

\usepackage[utf8]{inputenc}
\usepackage{color}
\usepackage[skins,theorems]{tcolorbox}

\usepackage{lipsum}

\usepackage{xr}
\usepackage{cleveref}

\begin{document}

\title{{Motility-Driven Viscoelastic Control of Tissue Morphology in Presomitic Mesoderm}}
%
\author{Sahil Islam}
\email{ph22resch01009@iith.ac.in}
\affiliation{Department of Physics, Indian Institute of Technology, Hyderabad, Telangana, India}

\author{Mohd. Suhail Rizvi}
\email{suhailr@bme.iith.ac.in}
\affiliation{Department of Biomedical Engineering, Indian Institute of Technology, Hyderabad, Telangana, India}

\author{Anupam Gupta}
\email{agupta@phy.iith.ac.in }
\affiliation{Department of Physics, Indian Institute of Technology, Hyderabad, Telangana, India}

\begin{abstract}

Embryonic tissues deform across broad spatial and temporal scales and relax stress through active rearrangements. A quantitative link between cell-scale activity, spatial forcing, and emergent tissue-scale mechanics remains incomplete. Here, we use a vertex-based tissue model with active force fluctuations to study how motility controls viscoelastic response. After validation against experimental presomitic mesoderm relaxation dynamics, we extract intrinsic mechanical timescales using stress relaxation and oscillatory shear. The model captures motility-dependent shifts between elastic and viscous behavior and the coexistence of fast relaxation with long-lived residual stress. When subjected to spatially patterned, temporally pulsed forcing, tissues behave as mechanical filters: long-wavelength inputs are accumulated, whereas short-wavelength, cell-scale perturbations are rapidly erased, largely independent of motility. Simulations with localized motility hotspots, motivated by spatially confined FGF signaling reported in vertebrate limb development, produce sustained protrusive tissue deformations consistent with experimentally observed early bud-like morphologies. Together, these results establish a minimal framework linking motility-driven activity to wavelength-selective mechanical memory and emergent tissue patterning.
\end{abstract}

\maketitle

\section{Introduction}

Embryonic tissues operate far from mechanical equilibrium, where active cellular processes and passive material responses are tightly intertwined \cite{heisenberg2013forces}. 
During development, tissues undergo large-scale deformation and remodeling \cite{mongeraFluidtosolidJammingTransition2018}. 
This remodeling occurs via active cellular rearrangements and passive mechanical resistance, largely governed by cytoskeletal mechanics and actomyosin contractility \cite{kimEmbryonicTissuesActive2021a, PRXLife.1.013004, petridou2019tissue, krajncFluidizationEpithelialSheets2018}.
These mechanical responses span a wide range of spatial and temporal scales, making tissues highly sensitive to the nature of the applied deformation \cite{fruleuxGrowthCouplesTemporal2024}. 
Understanding the physical principles of this mechanoregulation is essential not only for interpreting spatially heterogeneous mechanical measurements but also for predicting how tissues reorganize and adapt under developmental and perturbative conditions.

Towards this, advanced experimental approaches, such as ferrofluid microdroplet deformation \cite{serwaneVivoQuantificationSpatially2017a, mongeraFluidtosolidJammingTransition2018}, 
magnetic bead twisting, and micropipette aspiration \cite{campas2014quantifying}, 
have enabled direct estimation of tissue rheology. 
Using these works, the viscoelastic nature of different embryonic tissues
has been established, where elastic and viscous responses are dominant on shorter and longer timescales, respectively \cite{forgacsViscoelasticPropertiesLiving1998}. 
For example, in the presomitic mesoderm (PSM) in vertebrates, 
a spatial gradient of mechanical properties exists along the anterior–posterior (A–P) axis, with both tissue stiffness and viscosity increasing from the posterior to the anterior end. 
This static pattern results in a posterior-to-anterior gradient from a more fluid-like to a more solid-like state across the tissue \cite{mongeraFluidtosolidJammingTransition2018, kimEmbryonicTissuesActive2021a}. 
Importantly, even at a fixed position along this axis, the mechanical response depends strongly on the timescale of deformation- elastic at short time scales and viscous at large timescales \cite{mongeraFluidtosolidJammingTransition2018, serwaneVivoQuantificationSpatially2017a}.  
This intrinsic timescale-dependent behavior is a hallmark of tissue viscoelasticity in the developing embryo.

These tissue-scale viscoelasticity gradients in vertebrate PSM are accompanied by variations in spatial motility of the cells. Time-lapse microscopy shows that cells in the anterior region undergo weaker random spatial motion than those in the posterior region \cite{benazerafRandomCellMotility2010}. This motility gradient closely correlates with the spatial distribution of fibroblast growth factor (FGF), a central regulator of cell movement, whose expression level decreases from posterior to anterior \cite{benazerafRandomCellMotility2010,regev2022rectified}. The coexistence of spatially varying motility and mechanical properties suggests a tight coupling between active cellular processes and emergent tissue rheology, providing a potential mechanistic basis for how developing tissues deform, reorganize, and pattern under physiological conditions.

Another aspect of tissues with graded mechanical properties is how cells perceive and respond to dynamic forces.
During embryogenesis, for example, tissues experience intrinsic and extrinsic pulsatile cues across wide spatiotemporal scales. 
Intrinsically, key morphogenetic events are driven by periodic cellular activities: actomyosin pulses produce ratcheted apical constrictions \cite{sutherland2020pulsed}, 
pulsed contractions power dorsal closure\cite{solon2009pulsed}, 
and oscillatory gene expression patterns direct boundary formation during somitogenesis \cite{hubaud2014signalling}. 
Extrinsically, rhythmic forces from neighboring organs also exert widespread influence. 
Early heartbeats, initiated before functional circulation, generate pulsatile pressure waves that propagate through millimeter-scale tissues and can mechanically stimulate adjacent structures such as the presomitic mesoderm \cite{hornberger2007rhythm,watanabe2014regular,tyser2020first}. 
Meanwhile, cilia-driven left–right nodal flow creates oscillatory shear stress that not only breaks bilateral symmetry but may also deliver mechanical stimuli to nearby mesodermal regions \cite{shinohara2017cilia}. 
Collectively, these cyclic mechanical inputs
help locally modulate cell behavior and globally orchestrate tissue-level morphogenetic processes. 
Thus, mechanical heterogeneity within developing tissues may serve as a critical modulator of rhythmic force transmission, enabling the spatiotemporal coordination of morphogenesis.

Despite substantial experimental progress, a predictive framework that connects active cellular motility, tissue viscoelasticity, and the spatiotemporal structure of mechanical cues remains lacking. 
In embryonic tissues, where pronounced gradients exist in both mechanical properties and cellular activity, it remains unclear how these features jointly determine whether a dynamic mechanical input is filtered, propagated, or permanently encoded as a tissue-scale deformation.
Here, we bridge this gap using a motile cell–based vertex model \cite{honda2022vertex} to investigate the mechanical response of the presomitic mesoderm. 
We first establish that the model reproduces key experimental signatures, including the characteristic bulging and rounding dynamics of PSM explants in vitro. 
Leveraging this validated framework, we extract the viscoelastic timescales that govern tissue mechanics along the anterior–posterior axis and develop analytical predictions for deformation under spatially patterned, pulsatile forcing. 
We also perform simulations to recapitulate these predictions, directly linking cell-scale motility, emergent viscoelasticity, and large-scale morphogenetic shape change.

\section{Active Vertex Model}
We use the vertex model \cite{honda2022vertex, fletcher2014vertex, farhadifar2007influence}, a well-established framework for modeling epithelial tissues, to study the emergence of viscoelasticity and its regulation in embryonic PSM. In the vertex model, individual cells are represented as polygons forming a confluent mesh [Figure~\ref{fig:vertex}(a)]. The tissue's mechanical behavior is described in terms of an energy functional that accounts for deviations from target cell geometry and interfacial tensions:
\begin{equation}
U = \sum_{c=1}^{N_c} U_c = \sum_{c=1}^{N_c} \lambda (A_c - A_0)^2 + \beta P_c^2 + \gamma P_c.
\label{eqn:energy}
\end{equation}
Here, $N_c$ is the total number of cells, $A_c$ and $P_c$ are the instantaneous area and perimeter of cell $c$, and $A_0$ is the preferred cell area. The parameters $\lambda$, $\beta$, and $\gamma$ respectively control the resistance to area deformations (bulk elasticity), cortical contractility, and effective interfacial tension arising from adhesion and membrane tension. The motion of each vertex follows overdamped dynamics governed by:
\begin{equation}
\eta \dot{\mathbf{r}}_{v} = -\nabla_v U  + \boldsymbol{\xi}_{v}(t).
\label{eqn:force}
\end{equation}
where $\mathbf{r}_{v}$ is the position of vertex $v$ [Figure~\ref{fig:vertex}(a)], and $\eta$ is an effective friction coefficient that captures viscous resistance from the substrate and/or surrounding fluid. The term $\nabla_v U $ represents the gradient of the energy $U$ taken with respect to vertex position. The stochastic term $\boldsymbol{\xi}_{v}(t)$ accounts for active movement of cellular junctions and is modeled as Gaussian white noise with zero mean and amplitude proportional to a motility parameter $\mathcal{M}(\mathbf{r})$ :
\begin{equation}
\begin{aligned}
\langle {\xi}_{v,\alpha}(t) \rangle &= 0, \\
\langle {\xi}_{v,\alpha}(t) {\xi}_{v',\beta}(t') \rangle &= 2\mathcal{M}(\mathbf{r})\eta\delta_{vv'} \delta_{\alpha \beta} \delta(t - t').
\end{aligned}
\label{eqn:motility}
\end{equation}
where $\delta_{vv'}$ ensures that the noise is uncorrelated between different vertices and ${\xi}_{v,\alpha}$ is $x$ or $y$ components of $\boldsymbol{\xi}_{v}$. 
Together with Eq.~\ref{eqn:energy} and  \ref{eqn:force}, this formulation captures both the mechanical forces that maintain cellular equilibrium and the stochastic, non-equilibrium forces. These stochastic forces effectively represent cellular motility, originating from intracellular processes such as cytoskeletal turnover and actomyosin contractility, and should not be confused with thermal noise \cite{regev2022rectified}. These forces create random movement of cells, which in the context of PSM, parallels  fibroblast growth factor (FGF) signaling in coherence with experiments reported in \cite{benazerafRandomCellMotility2010,regev2022rectified} (see Section \ref{sec:bulging} for details). Importantly, these random forces at cell vertices drive similar cell–cell contact length fluctuations observed in experiments done on actively fluctuating epithelial junctions~\cite{kimEmbryonicTissuesActive2021a}. \\
In this model, we also take into account cellular junctional rearrangements, including T1 transitions, where neighboring cells exchange contacts (Figure \ref{fig:vertex}b), and T2 transitions, where cells below a critical size undergo extrusion and are replaced by a multicellular junction (Figure \ref{fig:vertex}c). Cell proliferation is not included in this model, as the simulated timescales are significantly shorter than those of cell division.
\section{Active vertex model reproduces experimental observation involving PSM explants}{\label{sec:bulging}}
\begin{figure*}
    \centering
    \includegraphics[width=\linewidth]{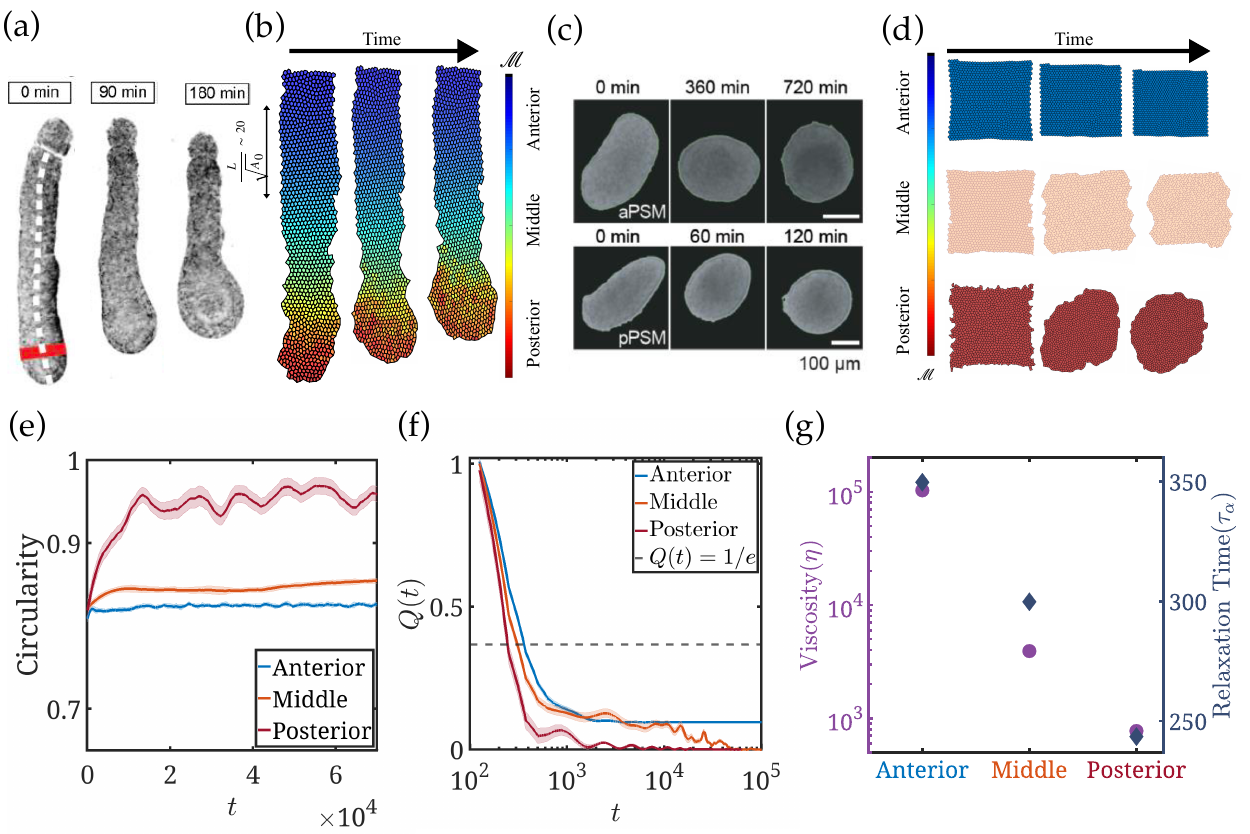}
    \caption{Effect of motility in presomitic mesoderm (PSM) rheology and relaxation. 
    (a) Change in the shape of the PSM explant over time. 
    Cell motility gradient drives pear-like shape formation of the explant (Experimental image is taken with permission from \cite{MICHAUT2025}).
    (b) A numerical simulation of the vertex model with a gradient in random cell motility in the form of exponential decay \cite{benazerafRandomCellMotility2010} from the posterior to anterior tissue has been introduced. 
    The model qualitatively reproduces the experimental observation shown in (a).
    (c) Examples of rounding of anterior and posterior explants from PSM (image taken with permission from \cite{MICHAUT2025}). The Anterior and posterior parts show significant distinction in rounding timescales, suggesting variation in viscosity along the A-P axis.
    (d) Numerical simulation of the vertex model of rounding of tissue from three different regions, anterior (blue), middle (green), and posterior (red) of the model PSM tissue.
    A higher cell motility results in faster rounding. 
    (e) Time evolution of circularity of the tissue with three different motility values (as in (d)).
    (f) Time evolution of overlap function $Q(t)$ (see Equation \ref{eqn:Qt}).
    It shows the rate of radial movement of a cell from its initial position. 
    The dotted line marks when $Q(t)$ drops below $1/e$ of its initial value.
    (g) Dependence of viscosity $\eta$, calculated from the Green-Kubo relation (see S.I. Sec.~\ref{sec:green-kubo} for definition), and $\alpha$ relaxation Time $\tau_{\alpha}$ on motility. The anterior region shows a high value of viscous timescales, which decreases towards the posterior direction.  
    }
    \label{fig:figure1}
\end{figure*}
To benchmark our simulation parameters ($\lambda$, $\beta$, $\gamma$) against PSM behavior, we reproduce an experimental result from \cite{MICHAUT2025}, where an embryonic PSM cultured ex vivo undergoes spontaneous shape remodeling. Experiments show that isolated PSM explants remodel their shape and fluidize over time \cite{MICHAUT2025}, demonstrating a tight coupling between motility and mechanical relaxation. By tuning our parameters to capture these behaviors, we establish a consistent framework for analyzing how spatial variations in motility and mechanical resistance along the anterior–posterior axis govern tissue morphogenesis.

\subsection{Cell motility drives bulging of PSM explant}
The PSM exhibits a graded profile of cellular motility along its anterior–posterior (AP) axis, regulated by fibroblast growth factor (FGF) signaling \cite{benazerafRandomCellMotility2010, regev2022rectified}. Highly motile cells in the posterior PSM can overcome intrinsic contractile and adhesive forces, enabling the explant to adopt a circular shape [Fig.~\ref{fig:figure1}(a)]. By contrast, the less motile anterior region fails to circularize, producing the characteristic pear-like morphology of the tissue [Fig.~\ref{fig:figure1}(a)]. This shape transformation occurs on timescales shorter than those of cell division \cite{MICHAUT2025}, indicating that it is driven by mechanics rather than growth.

To probe how mechanical properties vary along the AP axis, Arthur et al.\ \cite{MICHAUT2025} isolated tissues from different PSM regions and monitored their free evolution. Both anterior and posterior explants ultimately rounded into circular shapes, but on markedly different timescales [Fig.~\ref{fig:figure1}(c)]. Such rounding is a hallmark of fluidization, as a fluid minimizes surface energy by approaching a circular geometry. These observations demonstrate that motility-induced fluidization governs PSM morphogenesis.

We reproduce these experimental observations using an active vertex model. Starting from a rectangular tissue, we implement the FGF gradient by prescribing an exponentially decaying motility field, $\mathcal{M}(\mathbf{r})$, decreasing from posterior to anterior with a characteristic length scale set by experimentally measured motility decay \cite{benazerafRandomCellMotility2010, regev2022rectified, MICHAUT2025}. To remain consistent with experiments, we initialize the tissue in a state that initially contracts due to cellular contractility and adhesion. At longer times, motile cells overcome the energy barriers associated with junctional rearrangements, enabling fluidization. Because motility varies along the AP axis, fluidization occurs heterogeneously: the posterior PSM, with higher motility, rounds rapidly, whereas the anterior region remains elongated longer, ultimately producing the characteristic pear-like morphology accompanied by a net decrease in tissue length [Fig.~\ref{fig:figure1}(b)].

We further simulated tissue evolution under three distinct motility strengths. In agreement with experimental trends, tissues corresponding to posterior PSM reach circularity significantly faster than those representing middle or anterior regions [Fig.~\ref{fig:figure1}(d,e)].

In our simulations, tissue spreading emerges from motility-driven junctional rearrangements, particularly T1 transitions in which cells exchange neighbors. These rearrangements enable local structural reorganization. In regions with higher motility, such as the posterior PSM, cells more readily overcome the energetic barriers associated with junctional remodeling, leading to more fluid-like behavior (Fig.~\ref{fig:T1_count}). This spatially heterogeneous increase in tissue fluidity drives nonuniform deformation, reproducing the experimentally observed morphology. Our results thus demonstrate that active, fluctuation-driven rearrangements are sufficient to induce large-scale tissue remodeling even in the absence of cell proliferation.
  
\subsection{Motility gradient leads to rheological heterogeneity along PSM}

Experimental studies suggest that variations in tissue viscosity along the AP axis play a key role in shaping PSM morphogenesis \cite{mongeraFluidtosolidJammingTransition2018, MICHAUT2025}. To quantify this rheological heterogeneity in our simulations, we measured spatial variations in emergent tissue viscosity using two complementary approaches: the Green–Kubo formalism (S.I., Sec.~\ref{sec:green-kubo}) and the \textit{overlap function} method (S.I., Sec.~\ref{sec:overlap}).
The Green–Kubo analysis shows that the anterior PSM exhibits substantially higher viscosity than the posterior, with viscosity decreasing smoothly along the AP axis [Fig.~\ref{fig:figure1}(g)]. These results are in strong qualitative agreement with experimental measurements reported in ~\cite{MICHAUT2025}.
To further probe dynamical heterogeneity, we examined cell motions across three representative regions using the \textit{overlap function} $Q(t)$, which quantifies the radial displacement of cells from their initial positions. The resulting decay profiles display clear quantitative differences among anterior, middle, and posterior regions [Fig.~\ref{fig:figure1}(f)]. From these curves, we extracted the relaxation time $\tau_{\alpha}$ (see S.I. Sec.~\ref{sec:overlap}), which characterizes the timescale for structural rearrangements. We find that $\tau_{\alpha}$ is shortest in the posterior and longest in the anterior PSM [Fig.~\ref{fig:figure1}(g)], indicating that motility-driven fluidization is most pronounced toward the posterior. This gradient in viscous relaxation timescales confirms that tissue fluidity is spatially regulated and correlates with the observed morphological dynamics.


\section{Cell motility controls tissue viscoelasticity}

\begin{figure*}
    \centering
    \includegraphics[width=\linewidth]{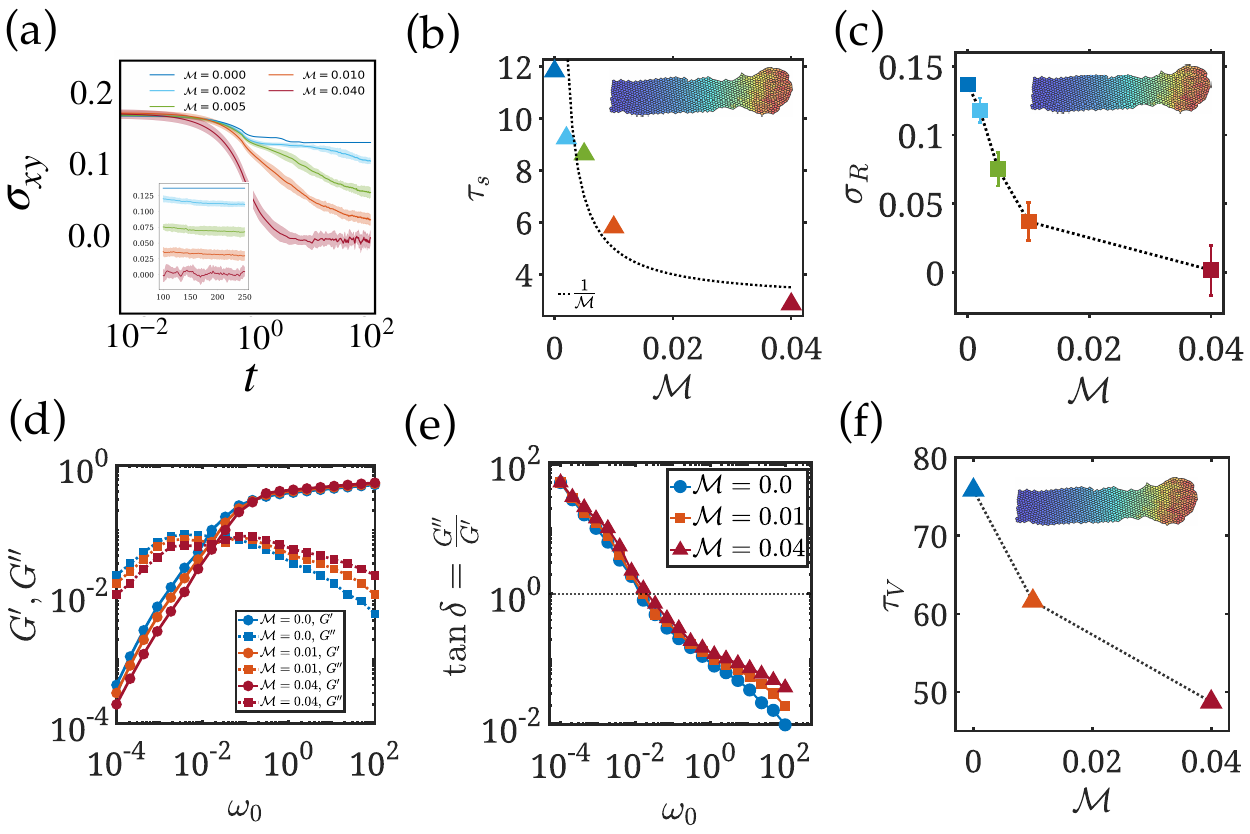}
  \caption{Tissue rheology under standard mechanical protocols.
(a) Stress relaxation response for tissues with different motility levels. The decay of shear stress \(\sigma_{xy}(t)\) is shown following a step shear deformation. Higher motility leads to faster stress relaxation and lower residual stress. \textit{Inset:} Long-time behavior of \(\sigma_{xy}\) reveals a nonzero plateau, indicating residual stress retention. 
(b) Relaxation timescale \(\tau_\mathrm{s}\) decreases with increasing motility, indicating enhanced fluidization. 
(c) Residual stress as a function of motility. More motile tissues retain less stress over time. 
(d) Frequency-dependent storage modulus \(G'\) and loss modulus \(G''\). At low frequencies, viscous behavior dominates (\(G'' > G'\)); at higher frequencies, the tissue responds more elastically. 
(e) The loss tangent \(\tan{\delta} = G''/G'\) as a function of driving frequency \(\omega_0\), illustrating the transition from viscous to elastic dominance. 
(f) Viscoelastic crossover time of the tissue($\tau_V$), defined by the crossover frequency where \(G' = G''\). $\tau_V$ shifts to lower values with increasing motility, reflecting faster stress relaxation dynamics in more active tissues.
}
    \label{fig:viscoelasticity}
\end{figure*}

To further probe tissue rheology, we subjected the tissue to controlled shear deformations under varying levels of cellular motility. This approach allowed us to quantify fundamental viscoelastic properties --- including stress-relaxation dynamics and the frequency-dependent storage and loss moduli --- and thereby establish a direct link between local active fluctuations and the emergent tissue-scale mechanics.

\subsection{Stress relaxation test reveals motility-driven residual stress retention}
The anterior–posterior variation in tissue viscosity suggests that different regions of the PSM may respond differently to mechanical loading. To investigate how this rheological heterogeneity affects stress dissipation, we performed stress-relaxation tests on tissues with varying motility levels. In each simulation, we applied a step shear by deforming the tissue affinely and fixing the outermost cell layers to maintain the applied strain. We then tracked the temporal evolution of shear stress in the bulk region, away from the boundary. By comparing tissues with low, medium, and high motility representing anterior, middle, and posterior PSM, we assessed how active cellular dynamics influence the tissue's ability to dissipate stress.

Across all motility regimes, we observed a characteristic biphasic relaxation response. At short to intermediate timescales, shear stress decayed rapidly [Fig.~\ref{fig:viscoelasticity}(a)], and the slope of this decay depended strongly on motility: higher motility produced steeper relaxation curves, indicating faster stress dissipation(Fig.~\ref{fig:viscoelasticity}(b)).

At longer timescales, the stress curves plateaued, revealing a non-zero residual stress in the tissue [Fig.~\ref{fig:viscoelasticity}(a), \textit{Inset}]. This behavior indicates solid-like features where cells cannot fully rearrange to relax all internal stresses. The magnitude of this residual stress was highest in the low-motility (anterior) tissues and decreased progressively with increasing motility (posterior) [Fig.~\ref{fig:viscoelasticity}(c)]. This motility-dependent decline in residual stress agrees qualitatively with experimental observations in embryonic PSM, where the posterior regions retain less residual stress than anterior ones~\cite{kimEmbryonicTissuesActive2021a, mongeraFluidtosolidJammingTransition2018}.

Together, these stress-relaxation results suggest that motility not only governs the rate of stress dissipation but also tunes the mechanical state of the tissue from a more elastic, stress-retaining regime to a more fluidized, actively relaxing state. These findings help explain how developing tissues adjust their mechanical behavior during morphogenesis by tuning how actively their cells move and rearrange.

\subsection{Oscillatory shear reveals cell motility modulates viscoelastic crossover frequency}

To probe the viscoelastic properties of the tissue, we subjected the model tissue to low-amplitude oscillatory shear across a wide range of frequencies. This approach enabled us to extract the frequency-dependent storage modulus ($G'$), representing the elastic (energy-storing) response, and the loss modulus ($ G''$), representing the viscous (energy-dissipating) response. Together, these moduli quantify the relative solid- and fluid-like behavior of the tissue over different deformation timescales.

At low oscillation frequencies, where the period of deformation exceeds the intrinsic stress-relaxation time of the tissue, we observed that $G'' > G'$ (Fig.~\ref{fig:viscoelasticity}(d)). This indicates a viscosity-dominated response, in which the tissue can fully relax the imposed stress before the next deformation cycle begins, making the tissue behave more like a fluid. This corresponds to situations where cells have sufficient time to undergo junctional rearrangements (T1 transitions), enabling tissue flow (Fig.~\ref{fig:viscoelasticity}(d)). 

In contrast, at high frequencies, the oscillation timescale becomes shorter than the tissue's internal relaxation time, resulting in an elastic-dominated response with $G' > G''$ [Fig.~\ref{fig:viscoelasticity}(d)]. Here, rapid shear cycle does not allow adequate time for cellular rearrangements or internal stress relaxation, causing the tissue to store rather than dissipate deformation energy.

This fluid-to-solid transition is further characterized by the loss tangent $\tan\delta = \frac{G''}{G'}$, which decreases monotonically with frequency [Fig.\ref{fig:viscoelasticity}(e)]. The characteristic crossover frequency($\omega_c$), defined by the condition $G''(\omega_c)=G'(\omega_c)$, marks the intrinsic relaxation timescale of the tissue. Interestingly, this crossover shifts to higher frequencies as cellular motility increases [Fig.\ref{fig:viscoelasticity}(f)], indicating that active cell motion accelerates stress relaxation and pushes the viscoelastic transition to faster timescales.

These results highlight the central role of active cellular processes in shaping the viscoelastic landscape of the tissue, tuning its mechanical state between fluid-like and solid-like regimes depending on both motility and deformation frequency.

\section{Temporally pulsated spatial perturbation generates permanent deformation in viscoelastic tissue}
\begin{figure*}
	\centering
	\includegraphics[width=\textwidth]{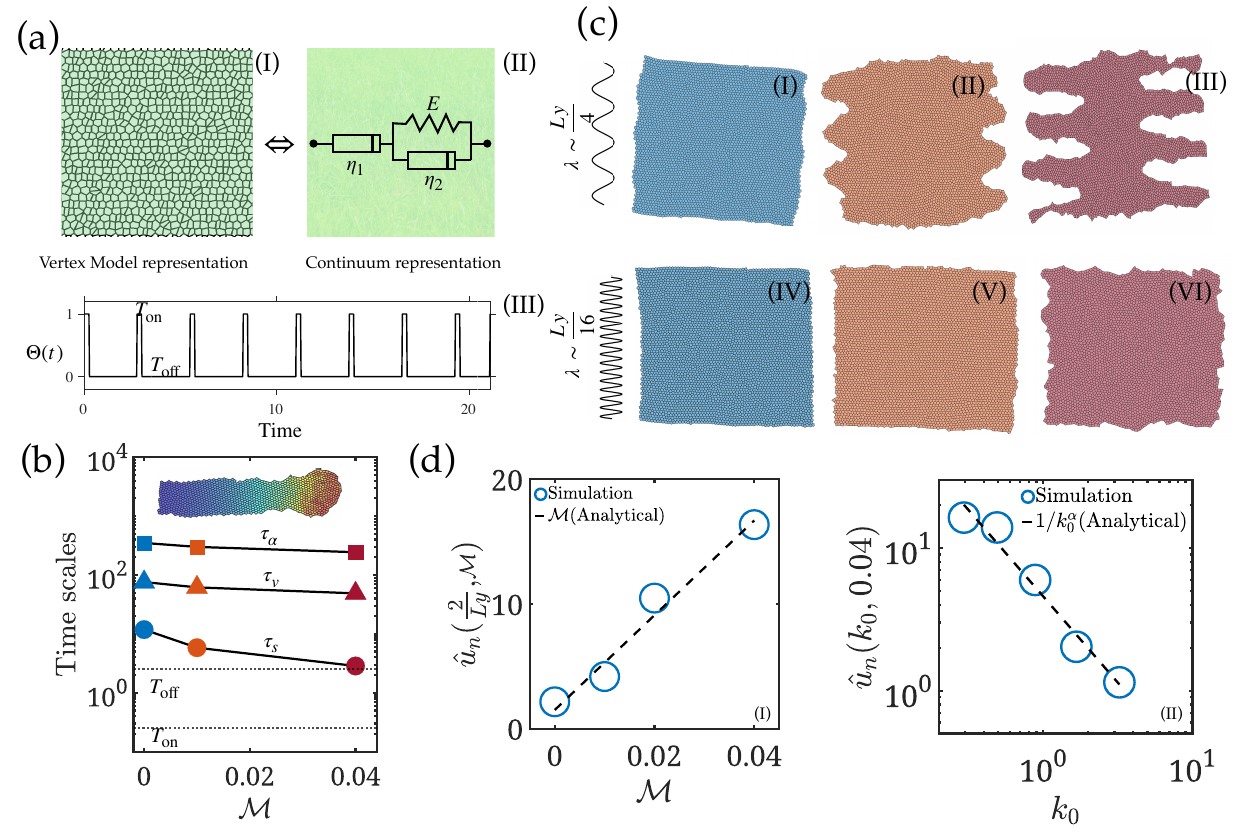}
	\caption{Tissue morphology under temporally pulsatile, spatially sinusoidal perturbations.  
		(a) Schematic of the analytical framework. (I) Synthetic tissue used from the vertex model. (II) Continuum representation of the tissue as a 2D viscoelastic material, modeled as a dashpot (viscosity $\eta_1$) in series with a Kelvin–Voigt element (dashpot with viscosity $\eta_2$ in parallel with a spring of elasticity $E$). (III) Generic time profile of the external forcing: the force remains on for a duration $\tauon$ and off for a period $\tauoff$.  
		(b) Comparison of characteristic timescales extracted from structural relaxation dynamics ($\tau_{\alpha}$) and standard rheological protocols ($\tau_s$ and $\tau_V$) along the anterior–posterior (A–P) axis of the PSM.  
		(c) Morphological outcomes of the vertex-model tissue under pulsatile forcing. (I) Low motility with long-wavelength perturbations ($\sim L_y/4$, where $L_y$ is tissue length along $y$) produces negligible morphological adaptation, but bulk rotation arises from force asymmetry. (II) Intermediate motility with the same wavelength induces moderate adaptation. (III) High motility yields pronounced morphological adaptation. (IV) Shorter-wavelength perturbations ($\sim L_y/16$) fail to elicit significant adaptation or rotation. (V, VI) No morphological adaptation is observed.  
		(d) Analytical predictions versus simulations of long-time tissue morphology in Fourier space, $\hat{u}_{n}(k_0, \mathcal{M})$, as a function of motility $\mathcal{M}$ and wavenumber $k_0$. Here $n$ is the number of on-off cycles. $n\gg1$ is chosen. (I) Predicted deformation increases linearly with motility, in quantitative agreement with simulations. (II) Theory predicts a power-law decay of deformation with increasing wavenumber, matching the simulation findings.}
	\label{fig:fig3_perturbation}
\end{figure*}

Having established the motility-dependent viscoelastic nature of the tissue, we next use the extracted viscoelastic timescales to predict tissue morphology under controlled external perturbations analytically.  Such analytical formulations offer a powerful means to uncover the fundamental principles governing complex tissue dynamics. By abstracting the system into a continuum viscoelastic framework, we can extract scaling relations, isolate key parameter dependencies, and obtain predictions that remain broadly generalizable across contexts.

\subsection{Analytical treatment shows motility-driven viscoelasticity and perturbation length scale controls permanent tissue deformation}

Motivated by the presence of multiple relaxation timescales in the tissue, we model the synthetic vertex tissue as a two-dimensional viscoelastic material using a linear viscoelastic model with two distinct relaxation timescales~\cite{kellyMechanicsLectureNotes2025}. This representation consists of a dashpot with viscosity $\eta_1$ in series with a Kelvin–Voigt element comprising a spring of modulus $E$ in parallel with a dashpot of viscosity $\eta_2$ [Fig.~\ref{fig:fig3_perturbation}(a)~(I), (II)]. The resulting constitutive relation between stress and strain is
	\begin{equation} \label{eqn:stress_strain}
		\tilde{\boldsymbol{\sigma}} + \frac{\eta_1 + \eta_2}{E} \, \dot{\tilde{\boldsymbol{\sigma}}}
		= \eta_1 \, \dot{\tilde{\boldsymbol{\varepsilon}}} + \frac{\eta_1 \eta_2}{E} \, \ddot{\tilde{\boldsymbol{\varepsilon}}},
	\end{equation}
where $\tilde{\boldsymbol{\sigma}}$ and $\tilde{\boldsymbol{\varepsilon}}$ denote the stress and strain tensors, respectively.  
We then subject this material to an external perturbation that is both spatially patterned and temporally pulsed. Specifically, the forcing has a sinusoidal spatial profile of wavelength $k_0$ and is applied in periodic on-off cycles of durations $\tauon$ and $\tauoff$:
	\begin{equation} 
		\begin{aligned}
			f(\mathbf{r},t) &= f_0 \, \sin (k_0 y) \, \Theta(t) ~\hat{\mathbf{x}}, \\[6pt]
			\Theta(t) &= \sum_{n=0}^{\infty} \mathrm{rect} \left( \frac{t - n(\tauon + \tauoff) - \tfrac{\tauon}{2}}{\tauon} \right), \\[6pt]
			\mathrm{rect}(\phi) &= 
			\begin{cases}
				1, & |\phi| \leq \tfrac{1}{2}, \\[6pt]
				0, & \text{otherwise},
			\end{cases}
			{\label{eqn:perturbation}}
		\end{aligned}
	\end{equation}
as illustrated in Fig.~\ref{fig:fig3_perturbation}(a)(III). Solving the constitutive equation under this forcing yields the deformation at the end of a single on–off cycle ($t=\tauon+\tauoff$), expressed in terms of the two characteristic relaxation times $\tau_1 = \eta_1/E$ and $\tau_2 = \eta_2/E$:
\begin{equation}
	\resizebox{\columnwidth}{!}{$ 
		\hat{u}_{t=\tauoff}(k_0, \tau_1, \tau_2,\tauon, \tauoff) = \frac{\tilde{f}_0}{E k_0^2}\left[\frac{\tauon}{\tau_1} + \frac{1}{E} \left(1 - e^{- \frac{\tauon}{\tau_2} } \right) e^{- \frac{\tauoff - \tauon}{\tau_2} }\right]
		$}
\end{equation}
(See S.I. Sec.~\ref{sec:analytical} for detailed calculations.)

From this, we can extract the dependence of the deformation after $n$ cycles, with one timescale $\tau_1$ and the perturbation wavelength $k_0$ as, 
\begin{equation}{\label{eqn:final_deformation_tau1}}
	\hat{u}_{n}(k_0,\tau_1) \sim \frac{n}{k_0^2 \tau_1}
\end{equation}

To connect this analytical deformation with experimentally accessible parameters such as motility, we identify $\tau_1$ with the stress-relaxation time, $\tau_s$. From Fig.~\ref{fig:viscoelasticity}(b), we established that $\tau_s$ scales inversely with motility, $\mathcal{M}$, i.e. $\tau_s \sim 1/\mathcal{M}$. Substituting this relation into Eq.~\ref{eqn:final_deformation_tau1} yields a compact scaling form for the steady-state deformation amplitude:
\begin{equation}{\label{eqn:final_deformation}}
	\hat{u}_{n}(k_0, \mathcal{M}) \sim \frac{\mathcal{M}}{k_0^2}.
\end{equation}
It is important to note that the analytical deformation, calculated at the end of a single $\tauoff$ interval, represents the elementary contribution per cycle. Over multiple cycles, this deformation accumulates to set the long-term behavior; however, the underlying scaling laws with $\mathcal{M}$ and $k_0$ remain unaffected. 

This scaling relation highlights several key features. First, for any finite $\tauon$, the material retains a nonzero residual deformation after each actuation cycle, enabling cumulative and potentially permanent shape changes.
Second, the deformation amplitude increases linearly with the inverse of the fast relaxation timescale $\tau_1$, indicating that faster-relaxing tissues respond more strongly to pulsatile forcing. Finally, Eq.~\ref{eqn:final_deformation} emphasizes the strong spatial dependence of the response: in the limit $k_0 \to \infty$ (short-wavelength perturbations), the deformation amplitude vanishes. Together, these results demonstrate that the interplay between cellular relaxation timescales and the spatial wavelength of external forcing jointly determines the efficiency of morphogenetic remodeling.

\subsection{Active vertex model tissue validates analytical prediction}
To test these analytical predictions, we subjected the synthetic tissue to a spatially patterned pulsed forcing identical in form to Eq.~\ref{eqn:perturbation}. Each cycle consisted of an active phase of duration $\tauon$ followed by a passive phase of duration $\tauoff$, chosen such that $\tauon < \tau_V \ll \tau_\alpha$ and $\tauoff \lesssim \tau_{\mathrm{s}}$ [Fig.~\ref{fig:fig3_perturbation}(b)]. These parameters were selected as an extreme regime relative to the analytical predictions to ensure that the simulated dynamics span well-separated timescales. This choice ensures that each active input pulse is delivered within the tissue's elastic response window, before complete stress relaxation. The forcing amplitude was kept below the intrinsic length scale of T1 transitions, thereby preventing the perturbation from directly inducing topological rearrangements. Consequently, the setup isolates how rapid, transient inputs interact with slower internal viscoelastic relaxation to generate long-term morphological outcomes. Despite the short-lived nature of each pulse, our simulations show that tissues can undergo significant morphological changes over time, depending on both cell motility and the spatial scale of the perturbation.

For long-wavelength perturbations ($\lambda \sim \frac{L_y}{4}, L_y$ is the length of the tissue along the y direction), tissue response exhibits a pronounced dependence on motility. In low motility regimes ($\mathcal{M} \sim 0.00$), the tissue stress relaxation timescale ($\tau_\mathrm{s}$) is much longer than the relaxation period of the applied perturbation ($\tauoff$) (Fig.~\ref{fig:fig3_perturbation}(b)). Consequently, stress fails to dissipate either via internal relaxation or via junctional rearrangements. Instead, the asymmetric and cyclic nature of the sinusoidal forcing drives a coherent bulk rotation of the tissue (a symmetric forcing eliminates bulk rotation), and no spatially structured patterns emerge [Fig.~\ref{fig:fig3_perturbation}(c)I]. At intermediate motility ($\mathcal{M} = 0.01$), isolated rearrangements occur, enabling partial imprinting of the pattern [Fig.~\ref{fig:fig3_perturbation}(c)II]. Only at high motility ($\mathcal{M} = 0.04$), where $\tau_\mathrm{s} \sim \tauoff$ (Fig.~\ref{fig:fig3_perturbation}(b)), tissue efficiently remodel via T1 transitions. In this regime, cells locally align with the imposed deformation, producing a clear and persistent morphogenetic pattern [Fig.~\ref{fig:fig3_perturbation}(c)III].

By contrast, short-wavelength perturbations ($\lambda \sim \tfrac{L_y}{16}$, roughly four cell diameters) fail to generate long-term morphological adaptation across all motility values. Unlike long-wavelength forcing, which imposes global deformations, short-wavelength perturbations generate localized edge-length variations arising from the rapid spatial alternation of applied forces. These sharp localized distortions trigger frequent T1 transitions, allowing the tissue to relax back to its initial state on timescales much shorter than those required for global deformation.

In the low-motility regime, one might expect slow stress relaxation to favor shape retention, as in the long-wavelength case. However, here the imposed short-wavelength fluctuations directly drive local junctional rearrangements, overriding the slow viscoelastic response and preventing any stable morphological adaptation [Fig.~\ref{fig:fig3_perturbation}(c)(IV)]. For intermediate motility, junctional rearrangements occur even more rapidly, further accelerating local relaxation and again erasing any imposed pattern [Fig.\ref{fig:fig3_perturbation}(c)(V)]. At the highest motility levels, cells actively remodel their junctions, and the combination of intrinsic motility with edge-length fluctuations leads to transient small-scale boundary undulations. Yet, unlike the long-wavelength perturbations that generate coherent and lasting morphological changes, these fluctuations are quickly dissipated and do not translate into stable adaptations [Fig.\ref{fig:fig3_perturbation}(c)(VI)].

Taken together, these results highlight a strong length-scale dependence in tissue mechanics. Long-wavelength perturbations couple to global viscoelastic modes, enabling robust, system-level adaptation, whereas short-wavelength perturbations are funneled into local junctional rearrangements and dissipated, leaving no lasting imprint.

Finally, we performed a quantitative comparison between analytical predictions and simulations by measuring the long-term deformation amplitude in Fourier space, $u_{n}(k_0, \mathcal{M})$ (see S.I. Sec.~\ref{sec:analytical}). The simulations show a linear increase in deformation amplitude with motility $\mathcal{M}$, in agreement with the analytical scaling (Fig.~\ref{fig:fig3_perturbation}(d)I). Both theory and simulations reveal a clear power-law dependence on perturbation length scale. While the theoretical prediction gives an exponent $\alpha = 2$, the simulations yield a slightly smaller value, $\alpha \approx 1.34$ [Fig.~\ref{fig:fig3_perturbation}(d)(II)]. Despite this quantitative difference, the underlying scaling behavior is preserved, validating the key analytical predictions.

\section{Discussion}
\begin{figure}
    \centering
    \includegraphics[width=\linewidth]{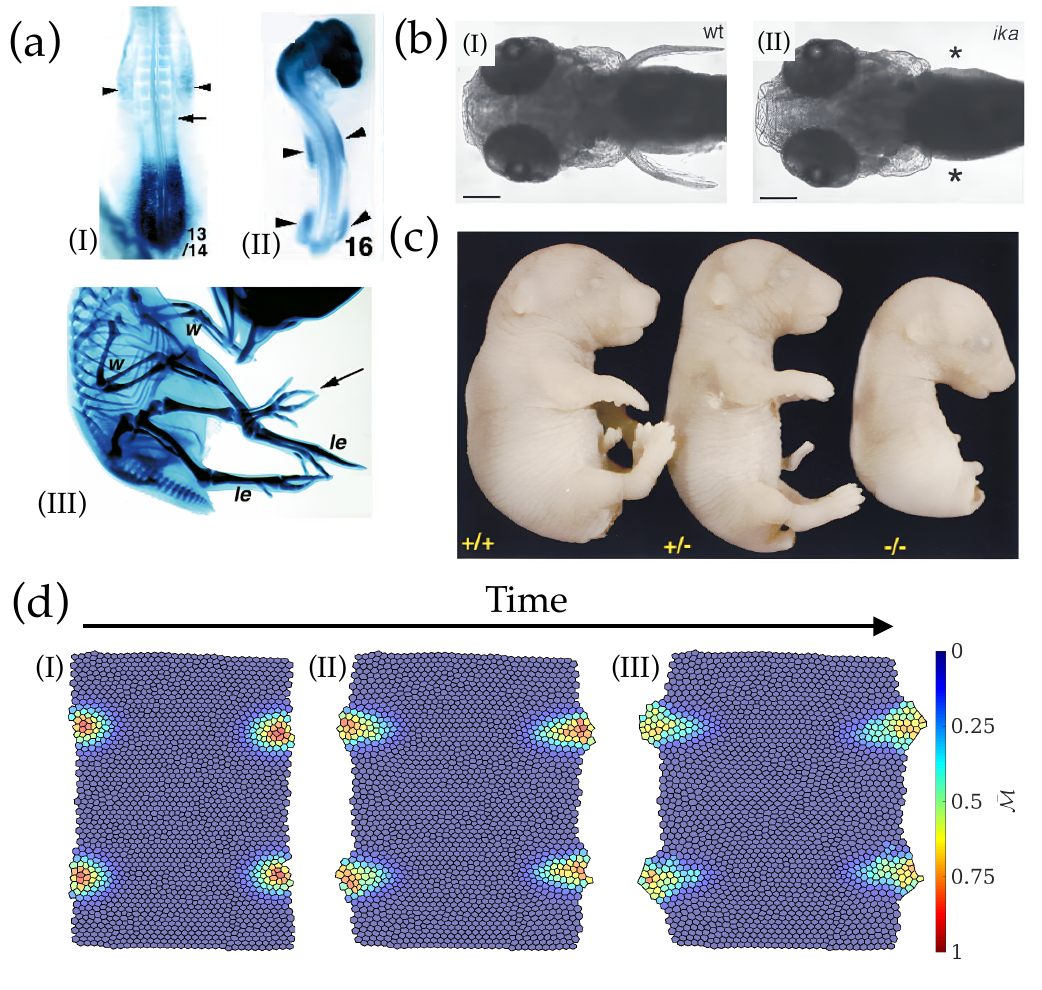}
    \caption{Effect of FGF signaling on vertebrate limb formation. (a) Chick embryo (adapted with permission from Ohuchi et al. \cite{ohuchiMesenchymalFactorFGF101997}): (I) Weak \textit{Fgf10} expression in the prospective forelimb mesoderm (arrowheads) is seen around 13 Hamburger and Hamilton (HH) stage. (II) \textit{Fgf10} expression in the head region and prospective limb mesoderm (arrowheads) increases at around the 16 HH stage. Suggesting \textit{Fgf10} plays a key role in limb bud formation. (III) Induction of an ectopic leg-like limb (arrow) following implantation of FGF10-expressing cells in the interlimb region. (b) Zebrafish embryo (adapted with permission from Fischer et al. \cite{fischerZebrafishFgf24Mutant2003}): (I) Wild-type larva at 3 dpf with pectoral fins protruding from the flanks. (II) \textit{ika} mutant (Fgf24-deficient) larva lacking pectoral fins (asterisks). (c) Mouse embryo (adapted with permission from Min et al.\cite{minFgf10RequiredBoth1998}): lateral views of \textit{Fgf10}$^{+/+}$, \textit{Fgf10}$^{+/-}$, and \textit{Fgf10}$^{-/-}$ embryos, showing complete limb absence in the knockout. (d) Simulations using active vertex model tissue to examine the role of FGF-induced motility in limb bud initiation. Starting from an initially compressed configuration, four localized motility hotspots (I) trigger protrusive outgrowths (II), which grow further resembling early limb bud formation (III). The colorbar represents motility value normalised with the maximum used for the posterior PSM.}
    \label{fig:LimbFormation}
\end{figure}
Our work provides a unified framework for understanding how cell motility modulates the viscoelastic response of embryonic tissues such as the presomitic mesoderm. Using an active vertex model calibrated to PSM explant relaxation measurements, we reproduced the motility-dependent mechanical behavior of PSM explants and uncovered how active fluctuations reshape tissue-level rheology. By characterizing the model under classical rheological protocols, such as stress relaxation and oscillatory shear, we identified the intrinsic mechanical timescales that govern morphogenetic patterning. These analyses show that motility accelerates stress relaxation by enhancing junctional rearrangements, effectively shifting the tissue from solid-like to fluid-like behavior. In low-motility regimes, where rearrangements are rare, residual stresses persist, and the tissue responds elastically over long timescales. In contrast, highly motile tissues rapidly reorganize to dissipate stress and adapt their morphology. This motility-dependent viscoelastic transition is consistent with in vivo observations in vertebrate PSM, where the posterior domain exhibits fluid-like behavior due to elevated cellular motility, while the anterior region remains more elastic \cite{serwaneVivoQuantificationSpatially2017a, mongeraFluidtosolidJammingTransition2018}.

A fallout of this motility-driven viscoelasticity is in embryonic tissues, which encounter rhythmic mechanical cues across a range of spatiotemporal scales, including periodic actomyosin contractions \cite{sutherland2020pulsed, solon2009pulsed}, oscillatory gene-expression rhythms during somitogenesis \cite{hubaud2014signalling}, early heartbeat-driven deformations \cite{hornberger2007rhythm, watanabe2014regular, tyser2020first}, and cilia-generated oscillatory flows \cite{shinohara2017cilia}. These cues have diverse origins but share a common physical role: they impose time-varying stresses that must be integrated by a viscoelastic, actively remodeling material.

To understand how such rhythmic cues sculpt tissue architecture, we developed a linear viscoelasticity-based analytical framework \cite{kellyMechanicsLectureNotes2025}. The model predicts that the tissue’s long-time deformation decreases sharply with increasing perturbation wavenumber $k_0$, reflecting a fundamental physical constraint: tissues cannot retain fine, cell-scale spatial features because these high–wavenumber inputs are rapidly dissipated through local junctional rearrangements. In effect, the tissue acts as a mechanical filter that selectively preserves long-wavelength, sub–tissue–scale cues while suppressing short-wavelength fluctuations. Guided by this principle, we examined the response of the active vertex tissue to pulsatile, spatially structured forcing. Even when each pulse was applied within the nominally elastic regime ($\tau_{\mathrm{on}} \ll \tau_V$), repeated cycles accumulated deformation in a motility-dependent manner—driving coherent alignment and pattern formation for long-wavelength inputs, but producing only transient, local distortions for short-wavelength perturbations. Consistent with the analytical predictions, highly motile tissues generated robust, persistent morphological patterns in response to long-wavelength cues, whereas cellular-scale perturbations were rapidly dissipated, leaving no lasting imprint.

Taken together, our results indicate that the spatial structure of an imposed cue and the local level of cellular motility jointly determine the bandwidth over which tissues retain and respond to biochemical or mechanical inputs. The intrinsic viscoelasticity of the tissue, therefore, governs how temporal and spatial features of external stimuli are integrated during morphogenesis. This framework provides a mechanistic bridge between molecular signaling and tissue mechanics: pathways such as FGF can modulate motility to tune viscoelastic responses, thereby shaping collective cell behaviors and morphogenetic flows.

One of the biological contexts where this connection between tissue rheology and external mechanical cues can play a major role is in the FGF signaling-driven initiation of the vertebrate limb bud.
In chick embryos, Fgf10 expression emerges in the presumptive forelimb mesoderm between HH13 and HH16, coinciding with the onset of limb-bud outgrowth \cite{ohuchiMesenchymalFactorFGF101997} (Fig.~\ref{fig:LimbFormation}(a) I,II). Experimental overexpression studies suggest that implanting Fgf10-producing cells in the interlimb region is sufficient to induce ectopic limb-like structures \cite{ohuchiMesenchymalFactorFGF101997} (Fig.~\ref{fig:LimbFormation}(a) III). This role is conserved across vertebrates: zebrafish fgf24 mutants fail to initiate pectoral fin outgrowth \cite{fischerZebrafishFgf24Mutant2003} (Fig.~\ref{fig:LimbFormation}(b)), and Fgf10$^{-/-}$ mice lack limb buds entirely \cite{minFgf10RequiredBoth1998} (Fig.~\ref{fig:LimbFormation}(c)). Collectively, these studies demonstrate that localized FGF activity provides a spatially confined signal that initiates bud formation.

To examine how such molecular cues could physically generate tissue-scale deformation, we used an active vertex model to simulate a mildly compressed epithelial sheet and introduced four spatially localized motility hotspots representing regions of elevated FGF activity. Under constrained extension along one axis, these hotspots produced persistent lateral cell outflow and drove a protrusive deformation reminiscent of early limb-bud emergence (Fig.~\ref{fig:LimbFormation}(d)). This minimal simulation illustrates how localized, FGF-driven increases in motility can convert molecular cues into emergent mechanical behaviors, offering a mechanistic interpretation that complements experimental observations across vertebrate systems.

Finally, the pulsatile-force analysis developed here has implications well beyond developmental morphogenesis. Many epithelial tissues in physiological settings experience rhythmic or cyclic mechanical loading. The alveolar epithelium, for example, undergoes continuous stretch–relax cycles during breathing and mechanical ventilation \cite{trepat2004viscoelasticity,tschumperlin2000deformation,geiger2009tubulin}. The intestinal epithelium is periodically deformed by peristaltic waves \cite{wang2018bioengineered}, and cardiac epithelia sustain high-frequency cyclic strain over billions of heartbeat cycles \cite{PAPAFILIPPOU2025102511}. Similar rhythmic deformations occur in the bladder, uterus, and other mechanically active organs. These contexts underscore that pulsatile inputs are not exceptional but represent a ubiquitous mode of mechanical forcing in living tissues.

Our framework shows that whether such cyclic cues accumulate into lasting morphological changes or are efficiently dissipated depends on the interplay between input frequency and the tissue’s motility-controlled relaxation dynamics. In settings such as mechanical ventilation, for instance, mismatches between driving frequency and intrinsic relaxation times may promote long-time deformation buildup, contributing to ventilator-induced injury. Conversely, adjusting deformation frequency or enhancing tissue fluidity—whether physiologically or through intervention—may reduce stress accumulation and preserve epithelial integrity. More broadly, our results indicate that tissues act as frequency-dependent mechanical filters, selectively retaining or erasing cyclic perturbations based on their viscoelastic state.

Overall, this work provides a general mechanistic framework for understanding how active, viscoelastic tissues integrate spatially and temporally structured mechanical cues. Beyond explaining observed behaviors in the PSM and limb initiation, the model offers a versatile platform for probing time-dependent pattern formation, stress encoding, and mechanical robustness in active biological materials. Extensions incorporating anisotropic tension generation, active nematic alignment, or mechano-chemical feedback will enable exploration of richer behaviors relevant to morphogenesis, organoid mechanics, and the engineering of programmable synthetic tissues.

\section{Acknowledgements}
These simulations were performed on the Paramseva supercomputers through the National Supercomputing Mission and the IITH Kanad clusters.
S.I. acknowledges the Prime Minister Research Fellowship (PMRF, ID-2002732) for financial support through a research fellowship. 
M.S.R. acknowledges SERB (India) project SRG/2021/001020 and IIT Hyderabad for financial support.  
A. G. acknowledges SERB-DST (India) Projects MTR/2022/000232, CRG/2023/007056-G, DST (India) grant no. DST/NSM/R\&D\_HPC\_Applications/2021/05 and grant no. SR/FST/PSI-215/2016, and IITH for Seed Grant No. IITH/2020/09.

\bibliography{Citations}

\clearpage
\setcounter{section}{0}
\setcounter{subsection}{0}
\setcounter{figure}{0}
\setcounter{table}{0}
\setcounter{equation}{0}
\renewcommand{\thesection}{S\arabic{section}}
\renewcommand{\thesubsection}{S\arabic{section}.\arabic{subsection}}
\renewcommand{\thefigure}{S\arabic{figure}}
\renewcommand{\thetable}{S\arabic{table}}
\renewcommand{\theequation}{S\arabic{equation}}

\clearpage

\onecolumngrid
\begin{center}
	{\Large\bfseries
		Motility-Driven Viscoelastic Control of Tissue Morphology in Presomitic Mesoderm\\
		(Supplementary Information)\par}
	\vspace{1em}
	
	Sahil Islam$^{1}$,
	Mohd.\ Suhail Rizvi$^{2}$,
	Anupam Gupta$^{1}$
	
	\vspace{0.5em}
	{\small
		$^{1}$Department of Physics, Indian Institute of Technology Hyderabad, India\\
		$^{2}$Department of Biomedical Engineering, Indian Institute of Technology Hyderabad, India
	}
\end{center}
\vspace{2em}

\twocolumngrid

\section{Methods}

\subsection{Active Vertex Model}

The energy functional for a single cell ($c$) in the vertex model is given by, 
\begin{eqnarray}
	U_c &= \lambda (A_c -A_0)^2 + \beta P_c^2 + \gamma P_c
\end{eqnarray}

\begin{figure*}
	\centering
	\includegraphics[width=0.9\textwidth]{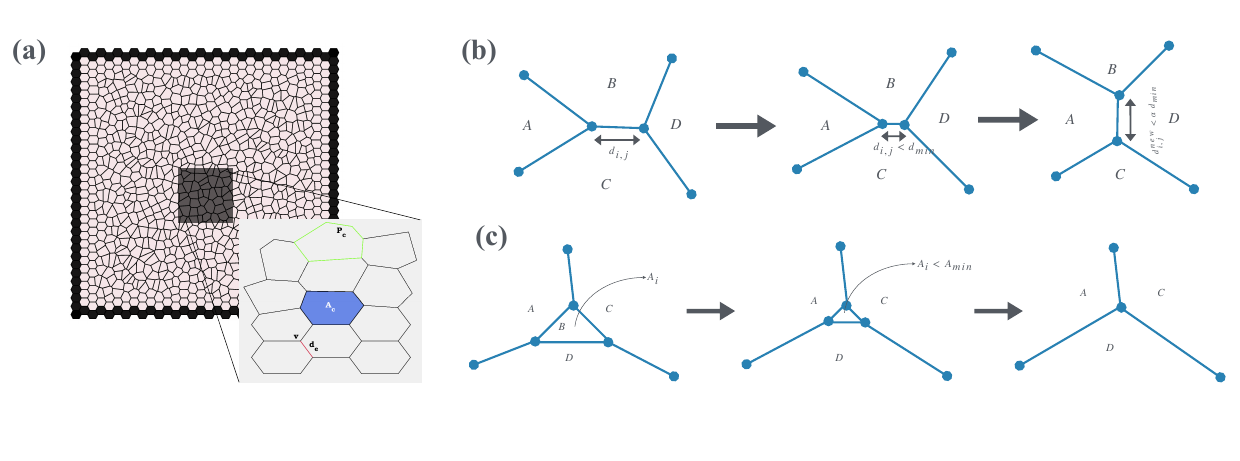}
	\caption{(a) A representative epithelial tissue configuration obtained from vertex model simulations. (b) Schematic illustration of a T1 transition involving neighbor exchange. (c) Schematic of a T2 transition representing cell extrusion.}
	\label{fig:vertex}
\end{figure*}

Where $A_c$ is the area and $P_c$ is the perimeter of a cell (Fig.~\ref{fig:vertex}(a)) and $A_0$ is the target area of a cell.  

The total force on a vertex ($v$) of a cell ($c$) is given as:

\begin{equation}
	\begin{aligned}
		\mathbf{F} 
		\renewcommand{\arraystretch}{1.5}
		= - \nabla_v(U_c) = 
		\begin{bmatrix}
			F_x \\
			F_y
		\end{bmatrix}=
		-2 \lambda (A_c - A_{0}) 
		\renewcommand{\arraystretch}{1.5}
		\begin{bmatrix}
			\frac{\partial Ac}{\partial x_v} \\
			\frac{\partial Ac}{\partial y_v}
		\end{bmatrix} \\
		- 2 \beta P_c
		\renewcommand{\arraystretch}{1.5}
		\begin{bmatrix}
			\frac{\partial P_c}{\partial x_v} \\
			\frac{\partial P_c}{\partial y_v}
		\end{bmatrix} 
		- \gamma 
		\renewcommand{\arraystretch}{1.5} 
		\begin{bmatrix}
			\frac{\partial P_c}{\partial x_v} \\
			\frac{\partial P_c}{\partial y_v}
		\end{bmatrix}.
	\end{aligned}
\end{equation}

The area of a polygon is given by:
\[
A_c = \frac{1}{2} \sum_{v=1}^{N_c} \left( x_v y_{v+1} - y_v x_{v+1} \right),
\]
where $N_c$ is the number of vertex in the cell $c$ and \(v+1\) is the next vertex index, and the indices are cyclic (i.e., $x_{N_c+1} = x_1$ ).

The perimeter of a polygon is given by:
\[
P_c = \sum_{v=1}^{N_c} \sqrt{\left( x_{v+1} - x_v \right)^2 + \left( y_{v+1} - y_v \right)^2}.
\]

Hence, the \(x\)- and \(y\)-gradients of \(A_c\) and \(P_c\) are:

\begin{equation}
	\frac{\partial A_c}{\partial \mathbf{r}_v} =
	\renewcommand{\arraystretch}{1.5}
	\begin{bmatrix}
		\frac{\partial A_c}{\partial x_v} \\
		\frac{\partial A_c}{\partial y_v}
	\end{bmatrix}
	=
	\frac{1}{2}
	\renewcommand{\arraystretch}{1.5}
	\begin{bmatrix}
		y_{v+1} - y_{v-1} \\
		x_{v-1} - x_{v+1}
	\end{bmatrix}
\end{equation}

\begin{equation}
	\frac{\partial P_c}{\partial \mathbf{r}_v} =
	\renewcommand{\arraystretch}{2}
	\begin{bmatrix}
		\frac{\partial P_c}{\partial x_v} \\
		\frac{\partial P_c}{\partial y_v}
	\end{bmatrix}
	=
	\renewcommand{\arraystretch}{2}
	\begin{bmatrix}
		\frac{x_v - x_{v-1}}{\sqrt{\Delta x_{v-1}^2 + \Delta y_{v-1}^2}} + \frac{x_v - x_{v+1}}{\sqrt{\Delta x_{v+1}^2 + \Delta y_{v+1}^2}} \\
		\frac{y_v - y_{v-1}}{\sqrt{\Delta x_{v-1}^2 + \Delta y_{v-1}^2}} + \frac{y_v - y_{v+1}}{\sqrt{\Delta x_{v+1}^2 + \Delta y_{v+1}^2}}
	\end{bmatrix}
\end{equation}

where \(\Delta x_{v-1} = x_v - x_{v-1}\), \(\Delta y_{v-1} = y_v - y_{v-1}\), and similarly for \(\Delta x_{v+1}\) and \(\Delta y_{v+1}\).

To make the cells motile we have added force as random white noise  $\xi_{v,\alpha}(t)$ with $\langle \xi_{v, \alpha}(t) \rangle  = 0 $  and  $\langle \xi_{v, \alpha }(t) \xi_{v',\alpha'}(t') \rangle  = 2 
\mathcal{M} (\mathbf{r}) \eta \delta_{v v'} \delta_{\alpha \alpha'} \delta(t-t') $ ) to each vertex position($v$) and $\alpha$ represents the direction of the force. 

\begin{equation}
	\eta \dot{\mathbf{r}}_{v} = -\nabla_v U  + \boldsymbol{\xi}_{v}(t).
	\label{eqn:vertex}
\end{equation}

This model incorporates cell junctional rearrangements (T1 transitions) and allows cells to detach from the tissue through T2 transitions (Fig.~\ref{fig:vertex}(b),(c)). 

The parameters involved in equation \ref{eqn:vertex} were non-dimensionalized using the length scale $L \sim \sqrt{A_0}$  and time scale $\frac{1}{\sqrt{\lambda A_0}}$. Both the values of $\lambda$ and $A_0$ were kept $1$ throughout all the simulations. 
The deterministic part of equation \ref{eqn:vertex} was solved by the implicit Euler time integration method, and the stochastic part was solved by the Wiener process increment with time step, $dt =  2.5\times10^{-3}$.

\section{Analysis}
\subsection{Geometric Properties}
\subsubsection{Tissue circularity}
\begin{figure}
	\centering
	\includegraphics[width=0.5\linewidth]{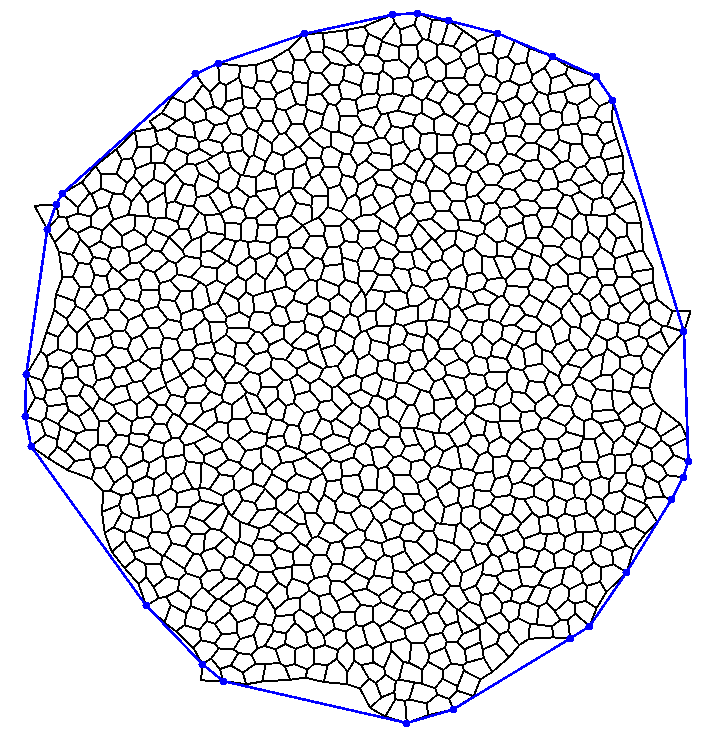}
	\caption{Using convex hull to evaluate the circularity of tissue morphology.}
	\label{fig:convexhull}
\end{figure}

\begin{figure}
	\centering
	\includegraphics[width=0.8\linewidth]{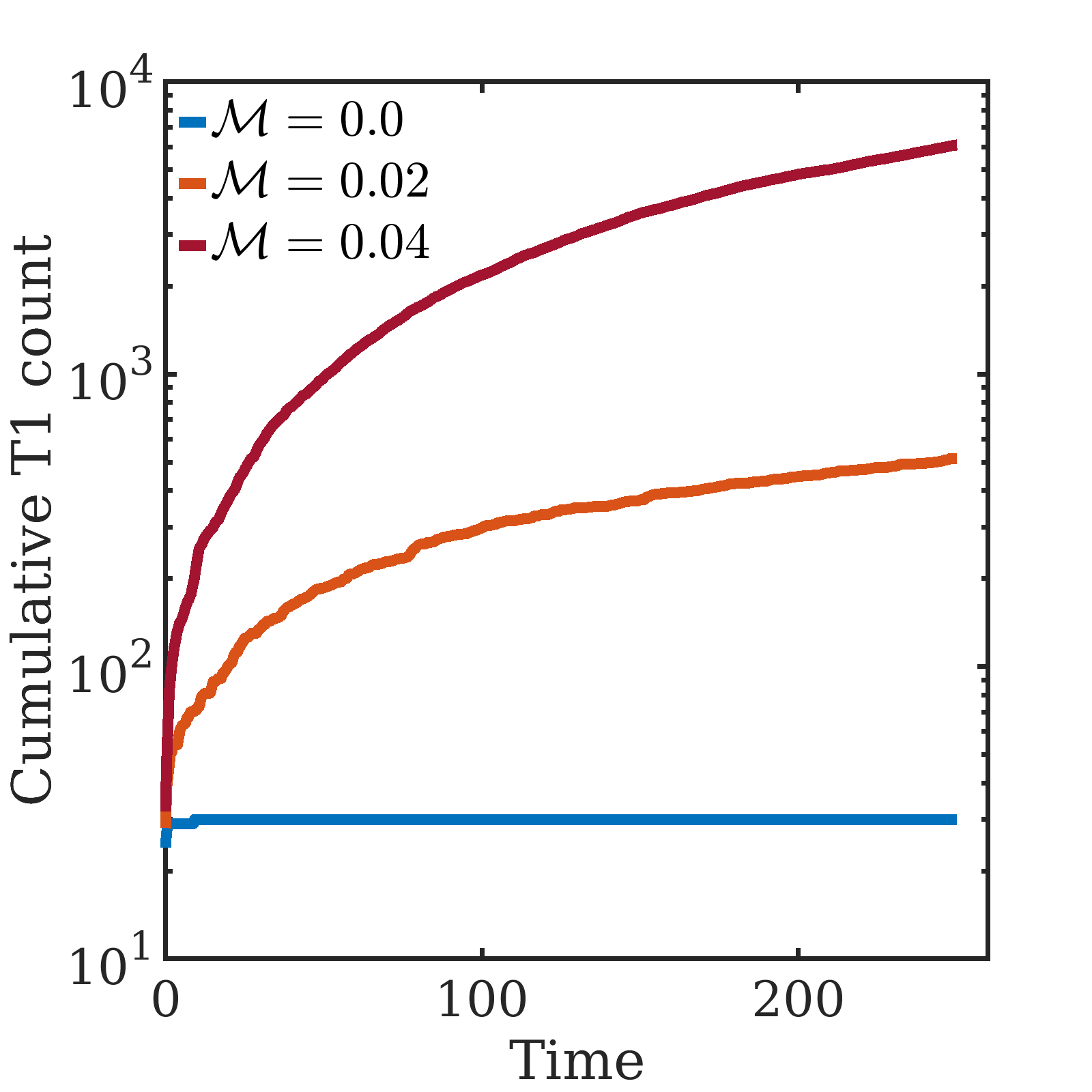}
	\caption{Cumulative count of T1 transitions ocer time. Higher motility promotes T1 transitions.}
	\label{fig:T1_count}
\end{figure}

We quantified tissue circularity by computing the convex hull of the final cell-center positions. From the resulting QHull polygon (Fig.~\ref{fig:convexhull}), we evaluated its area ($A$) and perimeter ($P$), and defined circularity as
\begin{equation}
	\text{Circularity} =  \frac{4 \pi A}{P^2}
\end{equation}
This gives a value of 1 for a perfect circle.

\subsection{Dynamic Properties}
\subsubsection{Overlap function} \label{sec:overlap}
The overlap function $Q(t)$ is defined by, 
\begin{equation}
	Q(t) = \left\langle \frac{1}{N_c}\sum_{c=1}^{N_c} W \left(a - |{\bf r} _c(t) - {\bf r}_c (0)| \right) \right\rangle
	\label{eqn:Qt}
\end{equation}
Where $W(x)$ is a Heaviside step function given by $W(x<0) = 1$. It quantifies the radial movement of a cell from its initial position. The relaxation time for a tissue is defined by the time it takes for $Q(t)$ to become $\frac{1}{e}^{th}$ of its initial value.  

\subsection{Rheological Properties}
\subsubsection{The stress tensor}
The stress tensor for an individual cell is calculated by, 
\begin{equation}
	\boldsymbol{\hat{\sigma}_c} = - \Pi_c \hat{\mathbf I} + \frac{1}{2 A_c} \sum_{e \in c} \boldsymbol{T}_e \otimes \boldsymbol{l}_e
\end{equation}
Where $ \Pi_c = - \frac{\partial U_c}{\partial A_c}$ is the hydrostatic pressure and $\boldsymbol{T}_e = \frac{\partial U_c }{\partial \boldsymbol{l}_e}$ is the line tension (or shear stress ) term. This leads to the expression, 
\begin{equation}
	\begin{aligned}
		\boldsymbol{\hat{\sigma}_c} =
		2 \lambda (A_c - A_0) 
		\begin{pmatrix}
			1 & 0 \\ 
			0 & 1
		\end{pmatrix} 
		+ \frac{1}{2 A_c} (2\beta P_c + \gamma ) \\
		\sum_{v = 1 }^{N_c} 
		\begin{pmatrix}
			\Delta x^2_{v+1} & \Delta x_{v+1} \Delta y_{v+1} \\
			\Delta y_{v+1} \Delta x_{v+1} & \Delta y^2_{v+1}
		\end{pmatrix}
		\frac{1}{\sqrt{\Delta x^2_{v+1} + \Delta y^2_{v+1}}}
	\end{aligned}
\end{equation}
To calculate the stress tensor for the whole tissue, we do a weighted sum $\sigma = \sum_{c=1}^{N_c} \frac{A_c}{A_{total}} \sigma_c$. 

\subsubsection{Green-Kubo Viscosity}{\label{sec:green-kubo}}
To calculate viscosity of the system as an emergent phenomenon, we have used Green-Kubo relation, 
\begin{equation}
	\eta = \frac{A}{\mathcal{M}} \int_0^\infty <\sigma_{\alpha\beta}(0)\sigma_{\alpha\beta}(t)> \rm{d} t
\end{equation}
Here $A$ is the area of the system and $\mathcal{M}$ is the motility of the system. The Green-Kubo method estimates the viscosity from the autocorrelation of shear stress.
\subsubsection{Stress relaxation experiment}
A stress relaxation experiment involves applying an external deformation to the system and observing how the internal stress dissipates over time. We apply an affine shear strain ${\bf \gamma}(t)=\left(\begin{array}{cc}1 & \epsilon \\ 0 & 1\end{array}\right)$ to the system and maintain the shear strain by keeping the boundary of the tissue fixed at a particular level of strain. We examine the relaxation of the shear stress of the bulk tissue, far away from the boundary. 
\subsubsection{Oscillatory shear}
We have applied oscillatory shear strain to the system with varying frequency of the oscillation and measured the stress response of the system. From this, we calculated Storage ($G'$) and Loss modulus ($G''$), given by the respective in-phase and out-of-phase responses of stress ($\sigma(t)$) to the applied strain ($\epsilon(t)$. 
\begin{equation}
	\begin{split}
		\epsilon(t) &= \epsilon_0 \sin( \omega_0 t)\\
		\sigma (t) &=\sigma_0 \sin (\omega_0 t+\delta) \\
		G' &= \frac{\sigma_0}{\epsilon_0} \cos(\delta) \\
		G'' &=  \frac{\sigma_0}{\epsilon_0} \sin(\delta)
		\label{eqn:oscl_shear}
	\end{split}
\end{equation}

\emph{Calculation of $G'$ and $G''$ from $\sigma(t)$ : } \\
From equation \ref{eqn:oscl_shear}, if we expand $\sigma(t)$ we get, 
\begin{equation}
	\begin{split}
		\sigma(t) &= \sigma_0 \left( \sin(\omega_0 t) \cos(\delta) + \cos(\omega_0 t) \sin(\delta) \right) \\
		\sigma(t) &= \epsilon_0 G' \sin(\omega_0 t) + \epsilon_0 G'' \cos(\omega_0 t))  
	\end{split}
\end{equation}
Thus, we have to find out the Fourier coefficients corresponding to the input frequency $\omega_0$ from the Fourier series of $\sigma(t)$. One way is to do a \emph{Fast Fourier Transform (FFT)} numerically and calculate the coefficients. However, this requires extensive sample data and can be prone to error at times. To avoid that, we have used an alternate calculation as follows: 
\begin{equation}
	f(t) = a_0 + \sum_{n=1}^{\infty} a_n \cos(\omega_n t) +  \sum_{n=1}^{\infty} b_n \sin(\omega_n t) 
	\label{eqn:fourier_trans}
\end{equation}
Let's say $n=n_0$ corresponds to the input frequency. So we have to find the coefficients of $\cos(\omega_{n_0} t)$ and $\sin(\omega_{n_0} t)$ i.e $a_{n_0}$ and $b_{n_0}$.
Now, if we integrate equation \ref{eqn:fourier_trans} to a time $T$ or multiply it with $\cos(\omega_{n_0} t)$ or $\sin(\omega_{n_0} t )$ and then integrate, we get three equations involving $a_0$, $a_{n_0}$ and $b_{n_0}$ as three unknowns which we can solve as system of linear equations. These equations look like,  
\begin{equation}
	\resizebox{\columnwidth}{!}{$
		\begin{bmatrix}
			\int_0^{T} f(t) \, dt \\
			\int_0^{T} f(t) \cos(\omega_{n_0} t) \, dt \\
			\int_0^{T} f(t) \sin(\omega_{n_0} t) \, dt
		\end{bmatrix}
		=
		\begin{bmatrix}
			\int_0^T 1 \, dt & \int_0^T \cos(\omega_{n_0} t) \, dt & \int_0^T \sin(\omega_{n_0} t) \, dt \\
			\int_0^T \cos(\omega_{n_0} t) \, dt & \int_0^T \cos^2(\omega_{n_0} t) \, dt & \int_0^T \sin(\omega_{n_0} t) \cos(\omega_{n_0} t) \, dt \\
			\int_0^T \sin(\omega_{n_0} t) \, dt & \int_0^T \sin(\omega_{n_0} t) \cos(\omega_{n_0} t) \, dt & \int_0^T \sin^2(\omega_{n_0} t) \, dt
		\end{bmatrix}
		\begin{bmatrix}
			a_0 \\
			a_{n_0}\\
			b_{n_0}
		\end{bmatrix}
		$}
	\label{eqn:matrix_eqn}
\end{equation}


In equation \ref{eqn:matrix_eqn}, the other terms where $n\ne n_0$ have been ignored as they would not contribute much. In equation \ref{eqn:matrix_eqn}, left-hand side integrals are calculated using a numerical integration scheme. 
Thus,  we can calculate $a_0$, $a_{n_0}$ and $b_{n_0}$, and from that we can get $G' = \frac{b_{n_0}}{\epsilon_0}$ and $G'' = \frac{a_{n_0}}{\epsilon_0}$. The amplitude of the oscillation was kept sufficiently small ($\epsilon_0\ll 1$) to probe a linear response. Simulations with higher motility required a higher ensemble average and low-pass filtering in the resulting stress before calculating the moduli.

\section{Analytical Treatment ( Viscoelastic material under pulsatile spatial perturbation )}{\label{sec:analytical}}

\begin{figure}
	\centering
	\includegraphics[width=\linewidth]{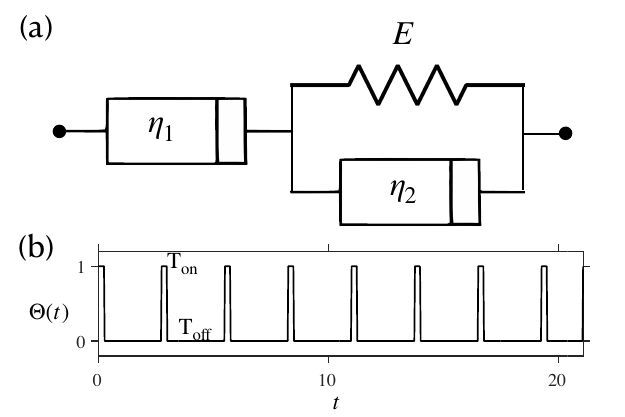}
	\caption{(a) Linear viscoelastic model with two timescales, consisting of a dashpot with viscosity $\eta_1$ in series with a Kelvin–Voigt element (spring with modulus $E$ and dashpot with viscosity $\eta_2$). (b) Time profile of the applied mechanical perturbation.}
	\label{fig:spring_dashpot}
\end{figure}

We investigated the response of a standard viscoelastic fluid subjected to spatially pulsatile perturbations, using a linear model with two timescales \cite{kellyMechanicsLectureNotes2025}. In this framework, a dashpot (viscosity $\eta_1$) is connected in series with a Kelvin–Voigt element, comprising a spring (modulus $E$) and a dashpot (viscosity $\eta_2$) in parallel(Fig~\ref{fig:spring_dashpot},(a)). The resulting stress–strain relationship is given by:

\begin{equation} \label{eqn:stress_strain}
	\tilde{\boldsymbol{\sigma}} + \frac{\eta_1 + \eta_2}{E} \, \dot{\tilde{\boldsymbol{\sigma}}}
	= \eta_1 \, \dot{\tilde{\boldsymbol{\varepsilon}}} + \frac{\eta_1 \eta_2}{E} \, \ddot{\tilde{\boldsymbol{\varepsilon}}}
\end{equation}

Where \( \tilde{\boldsymbol{\sigma}} \) is the stress tensor,
\( \tilde{\boldsymbol{\varepsilon}} = \frac{1}{2} \left( \nabla \mathbf{u} + \nabla \mathbf{u}^{\top} \right) \) is the linearized strain tensor,
\( \mathbf{u}(\mathbf{x}, t) \) is the displacement vector field,
\( E \) is the elastic modulus,
\( \eta_1 \) and \( \eta_2 \) are the viscosity. (Fig.~\ref{fig:spring_dashpot}(a)).

This material is subjected to a pulsated perturbation (Fig.~\ref{fig:spring_dashpot} (b)): 

\begin{equation}
	\Theta(t) = \sum_{n=0}^{\infty} 
	\operatorname{rect} \!\left( 
	\frac{t - n(\tauon + \tauoff) - \frac{\tauon}{2}}{\tauoff} 
	\right)
\end{equation}

\[
\operatorname{rect}(x) =
\begin{cases}
	1, & |x| \le \tfrac{1}{2}, \\[6pt]
	0, & |x| > \tfrac{1}{2}.
\end{cases}
\]

For $0 < t <  \tauon $, the forcing is 1, and for $\tauon < t < \tauoff$ the forcing is zero (Fig.~\ref{fig:spring_dashpot}(b)).

The force balance (quasistatic momentum conservation) is given by:

\begin{equation}
	\nabla \cdot \tilde{\boldsymbol{\sigma}}(\mathbf{x}, t) + \mathbf{f}_{\text{ext}}(\mathbf{x}, t) = 0,
\end{equation}

where \( \mathbf{f}_{\text{ext}} \) is the external body force per unit volume.

Taking divergence of Equation~\ref{eqn:stress_strain}, and defining,
\begin{equation}
	\begin{aligned}
		a &= \frac{\eta_1 \eta_2}{E}\\
		b &= \eta_1\\
		d &= \frac{\eta_1 + \eta_2}{E},
	\end{aligned}
\end{equation}

\begin{equation}{\label{eqn:force_strain}}
	-{\mathbf{f}}_{\text{ext}} - d \, \dot{{\mathbf{f}}}_{\text{ext}} =
	b \, \partial_t (\nabla \cdot \tilde{\boldsymbol{\varepsilon}})
	+ a \, \partial_t^2 (\nabla \cdot \tilde{\boldsymbol{\varepsilon}})
\end{equation}

We consider the general 2D displacement field:

\[
\mathbf{u}(x,y,t) = \begin{bmatrix} u_x(x,y,t) \\ u_y(x,y,t) \end{bmatrix}
\]

From here, 
\[
\nabla \mathbf{u} =
\begin{bmatrix}
	\partial_x u_x & \partial_y u_x \\
	\partial_x u_y & \partial_y u_y
\end{bmatrix}
\]

And the strain tensor looks like:
\[
\tilde{\boldsymbol{\varepsilon}} =
\begin{bmatrix}
	\partial_x u_x & \frac{1}{2}(\partial_y u_x + \partial_x u_y) \\
	\frac{1}{2}(\partial_y u_x + \partial_x u_y) & \partial_y u_y
\end{bmatrix}
\]

Hence, 

\[
\resizebox{\columnwidth}{!}{$
	- \begin{bmatrix}
		f_{\text{ext},x} \\
		f_{\text{ext},y}
	\end{bmatrix}
	- d \begin{bmatrix}
		\dot{f}_{\text{ext},x} \\
		\dot{f}_{\text{ext},y}
	\end{bmatrix}
	=
	\begin{bmatrix}
		b\, \partial_t \left(\partial_x^2 u_x + \frac{1}{2} (\partial_y^2 u_x + \partial_x \partial_y u_y) \right)
		+ a\, \partial_t^2 \left(\partial_x^2 u_x + \frac{1}{2} (\partial_y^2 u_x + \partial_x \partial_y u_y) \right) \\[1.5ex]
		b\, \partial_t \left(\partial_y^2 u_y + \frac{1}{2} (\partial_x^2 u_y + \partial_y \partial_x u_x) \right)
		+ a\, \partial_t^2 \left(\partial_y^2 u_y + \frac{1}{2} (\partial_x^2 u_y + \partial_y \partial_x u_x) \right)
	\end{bmatrix}
	$}
\]

Since the forcing is in the \(x\)-direction and varies only with y,  from the symmetry of the system, we make the ansatz for displacement field:
\[
u_x = u(y, t) \quad u_y = v(y,t),
\]

This implies:
\[
\partial_x u_x = 0, \quad \partial_x^2 u_x = 0, \quad \partial_y \partial_x u_y = 0
\]

The remaining non-zero derivatives are:
\[
\partial_y^2 u_x = \partial_y^2 u(y, t), \quad \partial_t \partial_y^2 u_x = \partial_t \partial_y^2 u(y, t)
\]

We know, $f_{\text{ext},x} = f_x$ and $f_{\text{ext},y} = 0$.\\\\
So, the force balance reads, 

\begin{equation}{\label{eqn:dir_deform}}
	\begin{aligned}
		-f_x - d \, \partial_t f_x &=
		\frac{b}{2} \, \partial_t \partial_y^2 u(y,t) +
		\frac{a}{2} \, \partial_t^2 \partial_y^2 u(y,t)\\
		0 &= b~\partial_t \partial_y^2 v(y,t) + b \partial_t \partial_y^2 v(y,t) 
	\end{aligned}
\end{equation}

We observe from Eqn~\ref{eqn:dir_deform} that the deformation in the y direction ($v(y,t)$) will decay with time.\\
To calculate the deformation in the x direction ($u(y,t)$), we need to solve Eqn~\ref{eqn:dir_deform}. We start by taking a Fourier transform to eliminate the spatial derivatives and solve the ODE in time in Fourier space. 

In Fourier space, the x-diraction force balance equation becomes (dropping the subscript x)
\begin{equation}\resizebox{\columnwidth}{!}{$
		\begin{aligned}{\label{eqn:ode_fourier}}
			-\hat{f}(k,t) - d \, \partial_t \hat{f}(k,t)
			&= -\frac{k^2}{2} \left( b \, \partial_t \hat{u}(k,t) + a \, \partial_t^2 \hat{u}(k,t) \right)\\
			\quad \implies \quad
			a \, \partial_t^2 \hat{u}(k,t) + b \, \partial_t \hat{u}(k,t)
			&= \frac{2}{k^2} \left( \hat{f}(k,t) + d \, \partial_t \hat{f}(k,t) \right)
		\end{aligned}
		$}
\end{equation}

\par
\par
Now, we have $f(\mathbf{r},t) =  f_0 \sin (k_0 y) \Theta(t)$. So, $ \hat{f}(k,t) = f_0  \cdot \pi \frac{1}{i} \left[ \delta(k + k_0) - \delta(k - k_0) \right] \Theta(t)$. \\\\
Hence, Eqn~\ref{eqn:ode_fourier} reads, 
\begin{equation}{\label{eqn:delta_function_form}} 
	\resizebox{\columnwidth}{!}{$ 
		a \, \partial_t^2 \hat{u}(k,t) + b \, \partial_t \hat{u}(k,t) = \frac{ f_o \pi}{i k^2} \left[ \delta(k + k_0) - \delta(k - k_0) \right] \left[ \Theta(t) + d \, \dot{\Theta}(t) \right]
		$} 
\end{equation}

We observe that the forcing term on the right-hand side of Equation~\ref{eqn:delta_function_form} is zero for any \(|k| \ne k_0 \). Since we are interested in the system's response to the external forcing, we restrict our attention to the mode \( k = k_0 \). Thus, we write Equation~\ref{eqn:ode_fourier} specifically for the \( k_0 \) mode as:

\begin{equation}{\label{eqn:u_Theta}}
	a \, \partial_t^2 \hat{u}(k_0,t) + b \, \partial_t \hat{u}(k_0,t)
	= \frac{\pi  f_0}{i k_0^2 } \left( \Theta(t) + d \, \dot{\Theta}(t) \right)
\end{equation}

Now, we are going to focus on the deformation in one cycle ($t = \tauon+\tauoff$), and also redefine the amplitude of forcing as complex amplitude $\tilde{f}_0 = \frac{\pi  f_0}{i}$. So, within one cycle, the equation Eqn~\ref{eqn:u_Theta} becomes, 

\begin{equation}{\label{eqn:main_deformation}}
	a \, \partial_t^2 \hat{u}(k_0,t) + b \, \partial_t \hat{u}(k_0,t)
	= \frac{\tilde{f}_0}{k_0^2} \left[ \theta(t) + d \left( \delta(t) - \delta(t - \tau_{\text{on}}) \right) \right]
\end{equation}
Here, $\theta(t) = 1 \text{ for } 0<t<\tauon$, and $\delta(t)$ is the dirac delta function. For notational simplicity we redefine, $c = \frac{\tilde{f}_0}{k_0^2}, d \equiv \frac{\tilde{f}_0}{k_0^2}d$.
So, we have, 
\begin{equation}{\label{eqn:main_deformation2}}
	a \, \partial_t^2 \hat{u}(k_0,t) + b \, \partial_t \hat{u}(k_0,t)
	=  c ~ \theta(t) + d ~ \left( \delta(t) - \delta(t - \tau_{\text{on}}) \right) 
\end{equation}

The boundary conditions for solving Eqn~\ref{eqn:main_deformation2} are, 

\begin{equation}{\label{eqn:boundary_cond}}
	\resizebox{\columnwidth}{!}{$ 
		\begin{aligned}
			\hat{u}(k_0, 0) &= 0, \quad 
			\partial_t \hat{u}(k_0, 0^+) = \frac{d}{a}, \\
			\hat{u}(k_0, \tauon^-) &= \hat{u}(k_0, \tauon^+), \quad
			\partial_t \hat{u}(k_0, \tauon^+) - \partial_t \hat{u}(k_0, \tauon^-) =  - \frac{d}{a}
		\end{aligned}
		$}
\end{equation}

For $0<t<\tauon$
\begin{equation}
	\begin{aligned}
		\hat{u}_\mathrm{on}(k_0, t) &= C_1 + \frac{a C_2}{b} e^{-\frac{b}{a} t} + \frac{c}{b} t \\
		\partial_t \hat{u}_\mathrm{on}(k_0, t) &= -C_2 e^{-\frac{b}{a} t} + \frac{c}{b}
	\end{aligned}
\end{equation}
And for $\tauon < t < \tauoff$
\begin{equation}
	\begin{aligned}
		\hat{u}_\mathrm{off}(k_0, t) &= C_3 + \frac{a C_4}{b} e^{-\frac{b}{a} (t-\tauon)}\\
		\partial_t \hat{u}_\mathrm{off}(k_0, t) &= -C_4 e^{-\frac{b}{a} (t-\tauon)}
	\end{aligned}
\end{equation}

After applying the boundary conditions (Eqn~\ref{eqn:boundary_cond}), the coefficients $C_1, C_2, C_3 \text{ and } C_4$ becomes, 

\begin{equation}
		\begin{aligned}
			C_2 &= \frac{c}{b} - \frac{d}{a} \\
			C_1 &= -\frac{a C_2}{b}\\
			C_4 &= C_2e^{-\frac{b}{a}\tauon} - \frac{c}{b} + \frac{d}{a}\\
			u_\mathrm{on}(\tau_\mathrm{on}) &= C_1 + \frac{aC_2}{b} e^{-\frac{b}{a}\tauon } + \frac{c \tauon}{b} \\
			C_3 &= u_\mathrm{on}(\tauon) - \frac{a C_4}{b}
		\end{aligned}
\end{equation}

Finally, with all the coefficients known, we can find the deformation at $t = \tauoff$, 

\begin{equation}{\label{eqn:final_u}}
	\hat{u}_{t=\tauoff}(k_0, a, b,C_3, C_4) = C_3 + \frac{a}{b} C_4 e^{-\frac{b}{a}(\tauoff - \tauon)}
\end{equation}

Since we are interested in the behavior of the final deformation with perturbation length scale and viscosity, after putting in the respective values of $C_1, C_2, C_3 \text{ and } C_4$ in Eqn~\ref{eqn:final_u}, we can get,

\begin{equation}
	\resizebox{\columnwidth}{!}{$ 
		\hat{u}_{t=\tauoff}(k_0, \tau_1, \tau_2, \tauon, \tauoff) = \frac{\tilde{f}_0}{E k_0^2}\left[\frac{\tauon E}{\eta_1} + \frac{1}{E} \left(1 - e^{- \frac{E}{\eta_2} \tauon} \right) e^{- \frac{E}{\eta_2} (\tauoff - \tauon)}\right]
		$}
\end{equation}

Identifying the two timescales $\tau_1 = \frac{\eta_1}{E}$ and $\tau_2 = \frac{\eta_2}{E}$, we rewrite the expression, as 
\begin{equation}
	\resizebox{\columnwidth}{!}{$ 
		\hat{u}_{t=\tauoff}(k_0, \tau_1, \tau_2, \tauon, \tauoff) = \frac{\tilde{f}_0}{E k_0^2}\left[\frac{\tauon}{\tau_1} + \frac{1}{E} \left(1 - e^{- \frac{\tauon}{\tau_2} } \right) e^{- \frac{\tauoff - \tauon}{\tau_2} }\right]
		$}
\end{equation}
From this, we can extract the dependence of the deformation
with one timescale $\tau_1$ and the perturbation wavelength
$k_0$ as,
\begin{equation}{\label{eqn:final_deformation}}
	\hat{u}_{t=\tauoff}(k_0, \tau_1) \sim \frac{1}{k_0^2 \tau_1}
\end{equation} 
Since this residual deformation accumulates in every cycle without affecting any scaling behavior, we can write the total deformation at the end of cycle $n$, 

\begin{equation}{\label{eqn:final_deformation}}
	\hat{u}_{n}(k_0, \tau_1) \sim \frac{n}{k_0^2 \tau_1}
\end{equation} 

To establish equivalence with our simulations, we identify the stress relaxation timescale $\tau_s$ from simulations with the viscous relaxation time $\tau_1$ in the linear viscoelastic model. Hence,

\begin{equation}{\label{eqn:final_deformation}}
	\hat{u}_{n}(k_0, \tau_1) \sim \frac{1}{k_0^2 \tau_s}
\end{equation}

Note, we drop the $n$ as it is a constant. 

Furthermore, since we have shown in the main text that motility $\mathcal{M}$ scales inversely with the stress relaxation time, i.e., $\mathcal{M} \sim 1/\tau_s$, we can re-express the deformation amplitude as:

\begin{equation}{\label{eqn:final_deformation}}
	\hat{u}_{n \gg 1 }(k_0, \mathcal{M}) \sim \frac{\mathcal{M}}{k_0^2}
\end{equation}

This highlights how increasing motility enhances the tissue's mechanical response to spatial perturbations.


\end{document}